\documentclass[conference]{IEEEtran}

\IEEEoverridecommandlockouts    

\usepackage{amsmath,amssymb,amsfonts}
\usepackage{algorithmic}
\usepackage{graphicx}
\usepackage{textcomp}
\usepackage{xcolor}
\usepackage[mathscr]{euscript}

\usepackage{cite}
\usepackage{tikz}
\usepackage{multirow}

\def\BibTeX{{\rm B\kern-.05em{\sc i\kern-.025em b}\kern-.08em
    T\kern-.1667em\lower.7ex\hbox{E}\kern-.125emX}}

\title{ Deep Q-Learning-based Distribution Network Reconfiguration for Reliability Improvement}

\author{Mukesh Gautam, \emph{Student Member, IEEE}, Narayan Bhusal, \emph{Student Member, IEEE}, \\ and Mohammed Benidris \emph{Senior Member, IEEE}, \\ Department of Electrical \& Biomedical Engineering, \\University of Nevada, Reno, Reno, NV 89557 \\
(emails: \{mukesh.gautam, bhusalnarayan62\}@nevada.unr.edu, and mbenidris@unr.edu)}

\begin{document}
\maketitle
\begin{abstract}
Distribution network reconfiguration (DNR) has proved to be an economical and effective way to improve the reliability of distribution systems. As optimal network configuration depends on system operating states (e.g., loads at each node), existing analytical and population-based approaches need to repeat the entire analysis and computation to find the optimal network configuration with a change in system operating states. Contrary to this, if properly trained, deep reinforcement learning (DRL)-based DNR can determine optimal or near-optimal configuration quickly even with changes in system states. In this paper, a Deep Q Learning-based framework is proposed for the optimal DNR to improve reliability of the system. An optimization problem is formulated with an objective function that minimizes the average curtailed power. Constraints of the optimization problem are radial topology constraint and all nodes traversing constraint. The distribution network is modeled as a graph and the optimal network configuration is determined by searching for an optimal spanning tree. The optimal spanning tree is the spanning tree with the minimum value of the average curtailed power. The effectiveness of the proposed framework is demonstrated through several case studies on 33-node and 69-node distribution test systems.
\end{abstract}
\begin{IEEEkeywords}
Deep Q Network, distribution system reliability, network reconfiguration, reinforcement learning, spanning tree.
\end{IEEEkeywords}

\section{Introduction}
The goal of electric utilities is to reliably and economically supply electricity to their customers through utilization of available resources. Since power interruptions are mostly due to failure of the distribution system components \cite{AChowdhury2009}, enhancing the reliability of distribution systems is inevitable to provide uninterruptible electric power supply to customers. Reliability of distribution systems can be enhanced in two ways: (a) optimal utilization of available resources using smart grid technologies and (b) installation of redundant resources. The option of installing redundant resources is not economical and waste of resources. Therefore, sophisticated smart grid technologies should be developed to optimally utilize distribution system resources in an optimum manner. In this context, distribution network reconfiguration (DNR) is one of such smart grid technologies to provide economic and reliable supply of electricity. DNR can optimize existing resources by modifying the configuration of distribution networks through changing status of sectionalizing and tie-switches.


Several analytical and population-based intelligent search approaches have been proposed in the literature to solve the DNR problem. 
A genetic algorithm-based DNR has been proposed in \cite{gupta2014distribution} for power quality and reliability improvement. Similarly, genetic algorithm has been used for DNR to improve the reliability and optimal placement of distributed generators in \cite{7764202}. In \cite{7858938}, mixed-integer quadratic programming has been used for reliability constrained power loss minimization. In \cite{zhang2012reliability}, neighbor search algorithm has been used for DNR to improve the reliability and reduce the loss by taking into account the uncertainties of data. A review on DNR approaches that improve the reliability and reduce the power loss is provided in \cite{sultana2016review}. Analytical and population-based intelligent search methods used for DNR to improve reliability has the following shortfalls. Accuracy and effectiveness of analytical-based methods for DNR depend upon the accuracy of models used, where accurate models impose scalability challenges. Also, mathematical models are usually derived based on several approximations and they require complete system information. Population based methods, on the other hand, are computationally expensive due to the large search space, especially when system sizes increase.

In addition to population-based intelligent search techniques and analytical methods, learning-driven approaches are gaining significant attention for optimal DNR. In \cite{gao2020batch}, batch-constrained reinforcement learning has been used for the dynamic DNR with the objective of minimizing network operational costs. In \cite{ji2021real}, network power loss and number of switching actions in distribution network are minimized using deep learning. Reinforcement learning has also been used to simultaneously reduce network power loss and improve voltage profiles \cite{9455432}, where loop-based encoding is leveraged with Noisy Net deep Q learning to improve training effectiveness and computational efficiency. 
In \cite{oh2020online}, deep Q learning has been implemented to minimize switching actions while performing the DNR. The work presented in \cite{oh2020online} has been tested for the computational cost and the scalability as compared to brute-force search algorithm and the genetic algorithm. Although there are several similarities between DNR for different objectives, optimum DNR for reliability improvement is a challenging task since it requires estimating energy not supplied for each possible configuration. Therefore, developing intelligent learning-based approaches for DNR to improve the reliability is pivotal.

This paper proposes a deep reinforcement learning (DRL)-based framework for DNR to improve the reliability of distribution systems. In the proposed optimization framework, the average curtailed power is used as an objective function. In addition to operation and technical constraints, all-node-traversing and radiality constraints are considered. In the training phase of the proposed framework, Q values are predicted using forward propagation of a deep neural network (DNN). Actions are selected using Epsilon-Greedy algorithm. When actions are passed through the training environment, the DRL agent gets rewarded (or penalized) based on its performance. Target Q values are calculated based on the reward. The mean squared error (MSE), which is the most commonly employed loss function for regression, is computed using the predicted and target Q values. Errors are then back-propagated to update the weights of DNN. The trained DRL agent is then used to find the best network configuration. The proposed framework is validated through case studies on several distribution test systems, and the results show that the proposed framework can effectively find a network configuration with high reliability level.

The main contributions of this paper are summarized as follows:  
\begin{itemize}
\item The proposed work represents an advancement over existing DNR approaches that either use approximate mathematical models or heuristic techniques. In contrary to existing analytical and population-based methods, learning-driven methods provide a flexibility for optimal DNR for changing system states (change of loads, addition of distributed energy sources, etc.) since they do not need to repeat the entire calculation processes when there is a change in system states. If trained properly, DRL can find the optimal or near-optimal network configuration in almost real-time.    
\item This papers provides a framework for optimal network reconfiguration with an objective of reliability enhancement. Formulating an objective function for reliability enhancement through DNR has been a challenging task because the the reliability level needs to be evaluated for each system configuration. 
\end{itemize}

The rest of the paper is arranged as follows. Section \ref{formulation} explains problem formulation for proposed DNR. Section \ref{proposed} describes the proposed framework and solution approach. Section \ref{cases} validates the proposed work through case studies. Section \ref{conclusion} provides concluding remarks.

\section{Problem Formulation}\label{formulation}
This section explains the formulation of the objective function and the constraints of the DNR problem.
\subsection{Objective Function}
Reliability is one of the major factors that indicates the performance of the system. Reliability of distribution systems can be quantified using several reliability indices including system average interruption frequency index (SAIFI), system average interruption duration index (SAIDI), customer average interruption duration index (CAIDI), and average curtailed power. The average curtailed power is taken as the objective function for the problem under consideration since it can capture the severity of the outages and is directly affected by the topology or configuration of a distribution network. Therefore, in this work, the objective of the DNR is to minimize the total annual energy not supplied by the system. Mathematically, the 
average curtailed power can be expressed as follows.
\begin{equation}
    \mbox{Average Curtailed Power} = \sum_{k\in \Omega_k} P_{d,k}U_k \mbox{,} \label{eqn:ENS}
\end{equation}
where $P_{d,k}$ is the power demand at node $k$; $\Omega_k$ is the set of nodes with power demand; and $U_k$ is the average annual power unavailability duration at node $k$, which can be defined as follows.
\begin{equation}
    U_k =\sum_{l\in \Omega_{lk}}\lambda_l r_l \mbox{,} \label{eqn:Uk}
\end{equation}
where $\lambda_l$ is the failure rate of branch (or edge) $l$; $r_l$ is the outage duration (or repair time) of branch (or edge) $l$; and $\Omega_{lk}$ is the set of branches (or edges) between substation node and node $k$.

\subsection{Constraints}
During the process of selection of the optimal configuration of the distribution network, two constraints (radiality constraint and all-node-traversing constraint) should always be satisfied.
\begin{itemize}
    \item Radiality constraint: Radiality constraint is always maintained in a distribution system in order to design the protection coordination schemes. Each candidate configuration should have a radial topology since most of the practical distribution systems do not have loop structure. 
    \item All-node-traversing constraint: A distribution system operator should always configure the network in such a way that all loads are supplied with power in non-contingent scenarios. Therefore, for each candidate configuration, all-node-traversing constraint should always be satisfied.
\end{itemize}

If we search for a spanning tree, both of the aforementioned constraints are satisfied. In order to search for the spanning tree, we have to represent the distribution network as an undirected graph $\mathscr{G}=(\mathscr{N},\mathscr{E})$,  where $\mathscr{N}$ is a set of nodes (or vertices) and $\mathscr{E}$ is a set of edges (or branches). For the graph, a node-branch incidence matrix can be constructed after satisfying all-node-traversing constraint. If the node-branch incidence matrix is full ranked, then the radiality constraint is satisfied.
\section{Proposed Framework}\label{proposed}
This work leverages recently advanced reinforcement learning techniques for DNR to improve reliability of distribution systems. This section provides a brief overview of Deep Q learning, reward function, and training attributes of the Deep Q learning. 

\subsection{Deep Q Learning}
A reinforcement learning (RL) is a branch of machine learning in which an \emph{agent} learns to take suitable \emph{actions} to maximize cumulative \emph{reward} it gets from an uncertain \emph{environment}. In general, an RL system consists of four main integrants: policy, reward, value functions, and environment model. An agent decides the action to be taken based on the policy. The policy maps states to actions. When the agent takes an action, it gets rewarded (or penalized). Value function calculates the expected value of cumulative reward that a agent gets when it follows a certain policy.
There are different algorithms for RL. The choice of an algorithm depends on many factors such as continuous/discrete nature of states, continuous/discrete action-space, etc. For the DNR problem under consideration, the action-space is discrete in nature, which makes Q-Learning a suitable candidate for the problem. However, a basic Q-Learning needs large sized look-up tables where state-action values are stored. To avoid the use of large sized look-up tables, a deep neural network (DNN) is used as an action-value function approximator. The addition of DNN in the basic Q-Learning make the framework a Deep Q Network (DQN).
The update rule for action-value function in Q-learning is defined as follows \cite{zai2020deep}.
\begin{equation}
\begin{aligned}
    Q(S_t,A_t) \leftarrow Q(S_t,A_t)+\alpha\times[R_{t+1}\\+\gamma\times \max\limits_a Q(S_{t+1},A_{t+1})-Q(S_t,A_t)]
\end{aligned}
\end{equation}
where $A_t$ and $S_t$ are the action and state of an agent at $t^{\text{th}}$ iteration;  $Q(S_t,A_t)$ is the action-value function at $t^{\text{th}}$ iteration; $Q(S_{t+1},A_{t+1})$ is the action-value function at $(t+1)^{\text{th}}$ iteration; $\alpha$ is the learning rate; and $\gamma$ is the reward discount factor.

Instead of iteratively updating the action-value function, the DNN is trained and the parameters of the action-value function are optimized to minimize the mean-squared error (MSE) loss function (i.e., regression loss function), which is expressed as follows \cite{cao2020reinforcement}.
\begin{equation}
L(\theta)=\mathbb{E}[(Q(S_{t}, A_{t}\vert \theta)-y_t)^{2}]
\label{eqn:loss_fun}
\end{equation}
where $\mathbb{E}$ denotes expectation operator; $\theta$ denotes the parameter of action-value function $Q(S_t,A_t)$; and $y_t$ denotes the target action-value function, which is defined a follows.
\begin{equation}
    y_t = R(S_{t}, A_{t})-\gamma \times \max\limits_{a}Q(S_{t+1}, A_{t+1}\vert \theta^{\prime}) \label{eqn:targetQ}
\end{equation}
In \eqref{eqn:targetQ}, $R(S_t,A_t)$ denotes reward function at $t^{\text{th}}$ iteration and $\theta^{\prime}$ denotes the parameter of action-value function $Q(S_{t+1},A_{t+1})$. 

For the DNR problem under consideration, the status of sectionalizing and tie switches denotes the state; and the indices of opened branches (or edges) denote the actions.
\subsection{Reward Function}

A reward function based on average curtailed power is designed to evaluate the actions taken by the agent. For each action (here, set of edges to be opened), firstly, all-node-traversing constraint is checked. If this constraint is not satisfied, a high negative reward (or penalty) is given to the agent. After this, a node-branch incidence matrix is computed for the network configuration. If the node-branch incidence matrix is full-ranked, this means that the network is radial. Therefore, if radiality constraint is not satisfied, a high negative reward (or penalty) is given to the agent. After both the constraints are satisfied, the average curtailed power of the network configuration is calculated using annual power unavailability duration ($U_k$) and power demand ($P_{d,k}$) at each node $k$.

\subsection{Training Attributes}
The training of DQN is performed for a certain number of episodes ($n_{ep}$). The initial state of the system is the state with tie-switches open. The weights of DNN are initialized with some random values. In each episode, the predicted Q values corresponding to each edge of the system is computed based on forward propagation of DNN. For the selection of actions, the Epsilon-Greedy (exploration-exploitation) algorithm \cite{sutton2018reinforcement} is used. The value of exploration rate i.e., epsilon ($\varepsilon$) is initialized at 1. The epsilon is updated after each episode as follows.
\begin{equation}
    \varepsilon_{new} = \varepsilon_{old} - \frac{\varepsilon_{old}-\varepsilon_{min}}{n_{ep}}
\end{equation}
where $\varepsilon_{min}$ is the minimum exploration rate.
The target Q value of the DQN is computed using \eqref{eqn:targetQ}.
The MSE losses for each episode are computed based on \eqref{eqn:loss_fun} using the actual and target Q-values. These MSE losses are back-propagated to update the weights of the DNN.
\begin{figure}
    \centering
    \vspace{-1ex}
    \includegraphics[scale=0.7]{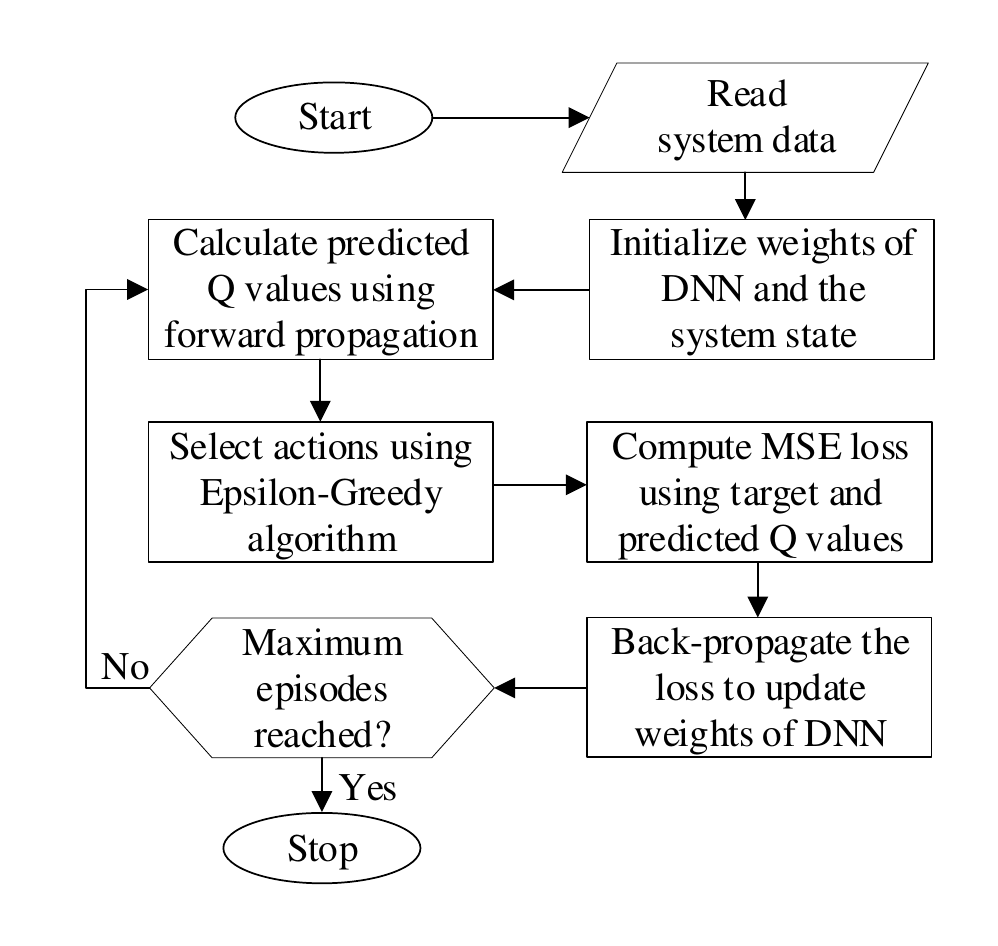}
    \vspace{-2ex}
    \caption{Flowchart showing the training phase of the proposed framework}
     \vspace{-1ex}
    \label{fig:flow_chart}
\end{figure}

The flowchart shown in Fig.~\ref{fig:flow_chart} describes the training phase of the proposed framework. 
\section{Case studies and Discussions}\label{cases}
The proposed framework is implemented on the $33$-node and $69$-node distribution systems. The case study parameters for both systems are as follows. The failure rate of each branch is assumed proportional to its branch impedance. The highest failure rate (in this case 0.4 failures/year or f/yr) is assigned to the branch with the largest impedance; the lowest failure rate (in this case 0.1 f/yr) is assigned to the branch with the smallest impedance; and linear interpolation is used to determine the failure rates of remaining branches. Regarding the outage duration (or repair rate) of each branch, its value is assumed to be constant (6 hr is used for all case studies). Normally-open switches (or tie switches) are assumed to be fully reliable. The annual power unavailability duration ($U_k$) of each node is determined based on the values of failure and repair rates for each branch.
\subsection{Case I: 33-node Distribution System}
 The $33$-node distribution test system is $100$ kVA, $12.66$ kV radial distribution system with $33$ nodes, $32$ branches and $5$ tie-lines. Therefore, the total number of branches in this system is $37$. All branches (including tie-lines) are numbered from $1$ to $37$ following our previous work \cite{gautam2020spanning}. The total load of the system is 3.71 MW. The detailed data of the system is provided in \cite{baran1989network}.
 \begin{figure}
    \centering
    \vspace{-1ex}
    \includegraphics[scale=0.55]{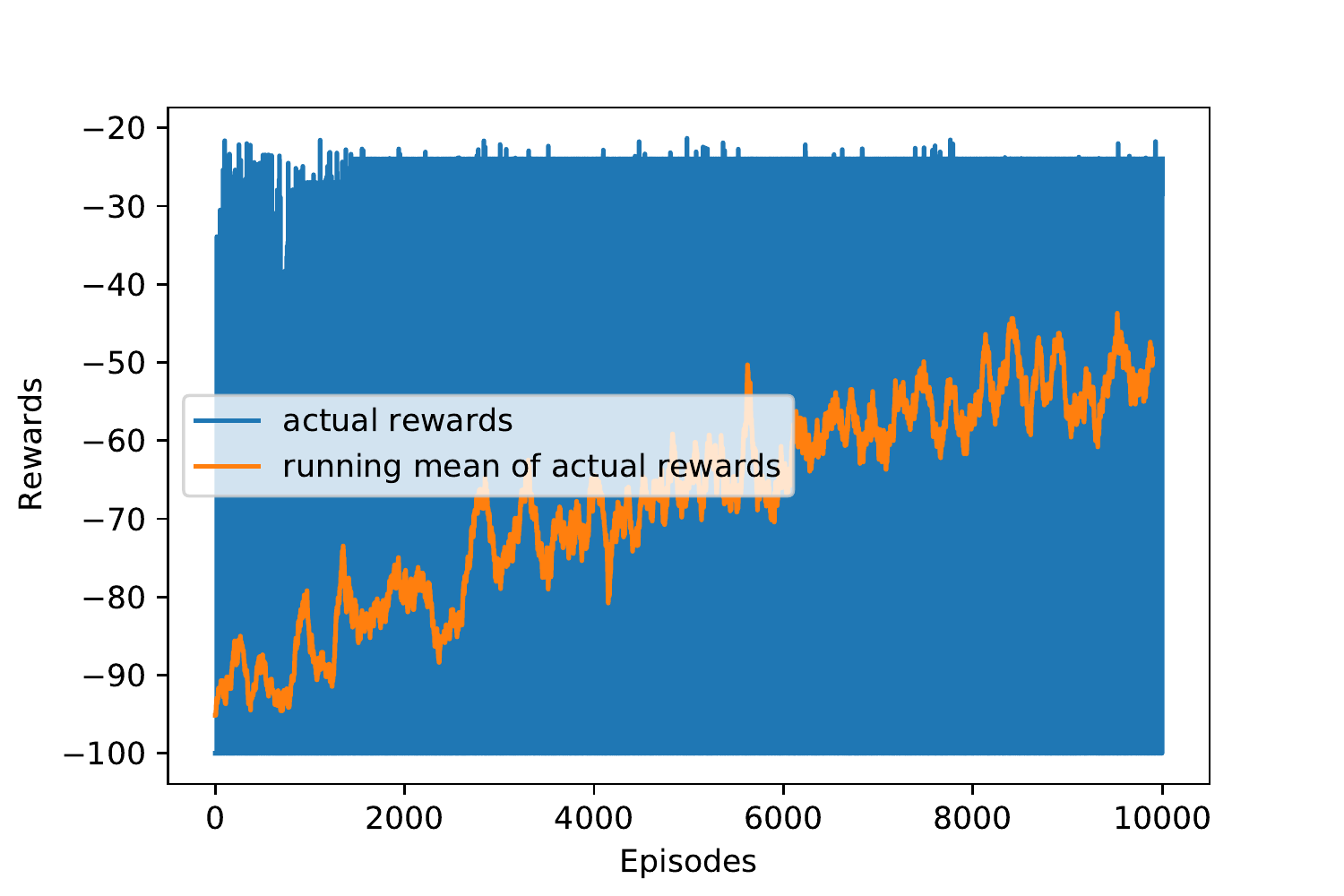}
    \vspace{-1ex}
    \caption{Rewards of training episodes for the 33-node system}
    \vspace{-1ex}
    \label{fig:33_rewards}
\end{figure}
\begin{figure}
    \centering
    \vspace{-1ex}
    \includegraphics[scale=0.55]{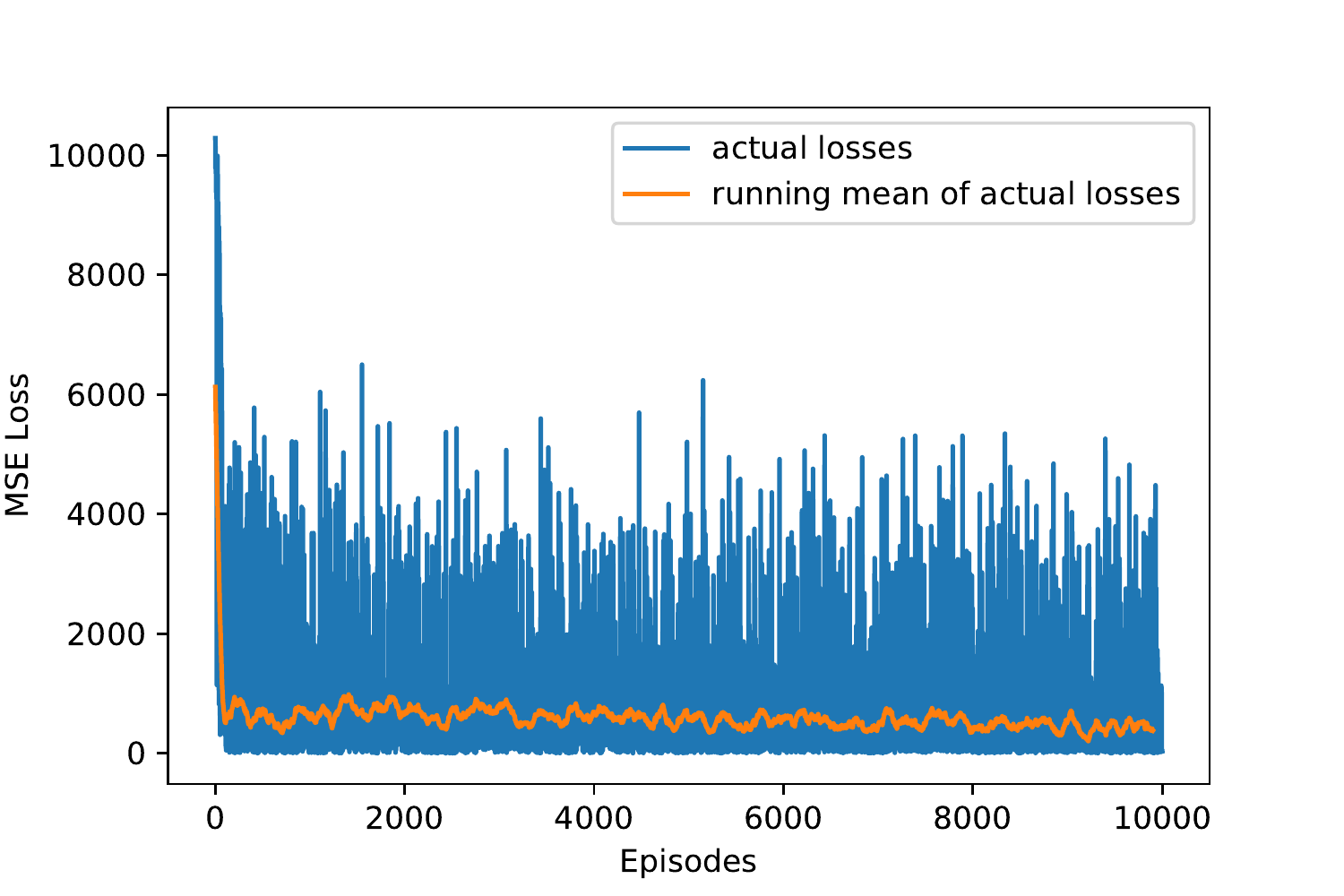}
    \vspace{-1ex}
    \caption{MSE of losses of training episodes for the 33-node system}
    \vspace{-1ex}
    \label{fig:33_losses}
\end{figure}
\begin{figure}
    \centering
    \includegraphics[scale=0.55]{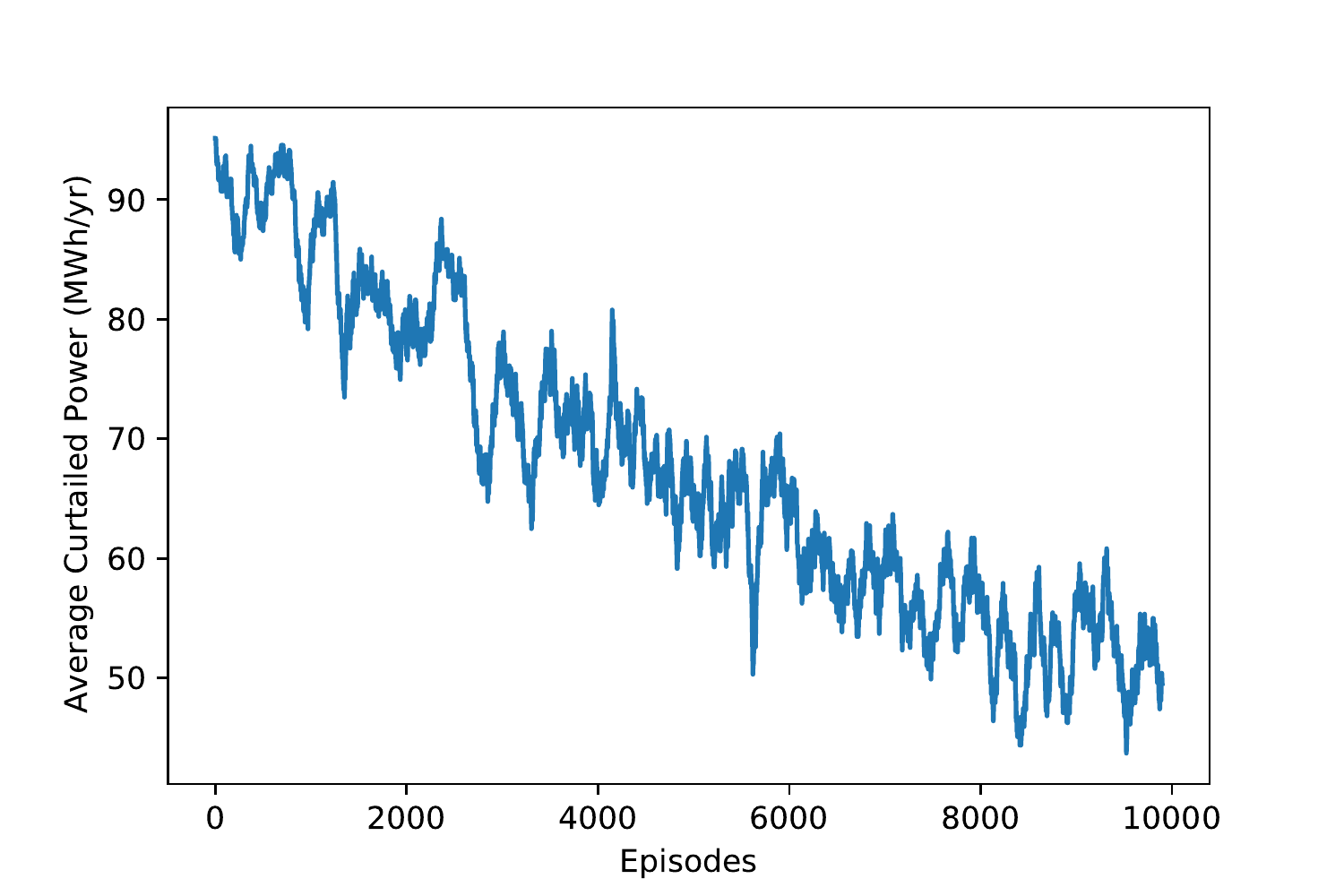}
    \caption{Average curtailed power of training episodes for the 33-node system}
    \label{fig:33_ENS}
\end{figure}

The training of the DQN for the 33-node system is performed for 10,000 episodes. The initial state of the system is the state with all tie-switches (i.e., edges 33, 34, 35, 36, and 37) are open. Initially, the rewards are very low but then they increase as the number of episodes increases. Fig.~\ref{fig:33_rewards} shows the actual rewards and running mean (100-episode window) of actual rewards as the episode progresses. It can be seen from Fig.~\ref{fig:33_rewards} that as the number of episodes increases, the running mean of the reward increases and saturates after nearly 8,000 episodes.

Fig.~\ref{fig:33_losses} shows the actual values and running mean (100-episode window) of MSE losses as the episode progresses. Fig.~\ref{fig:33_losses} also shows that initially MSE loss is very high but its running mean remains low after a few episodes. Fig.~\ref{fig:33_ENS} shows the plot of running mean (100-episode window) of the average curtailed power as the episode progresses. From Fig.~\ref{fig:33_ENS} we can see that the average curtailed power decreases as the number of episodes increases and settles down after nearly 8,000 episodes. The final value of the average curtailed power obtained during the training of the proposed framework is 23.96 MWh/yr. For this value of the average curtailed power, the open edges (branches) are 7, 14, 26, 33, and 34.

\subsection{Case II: 69-node Distribution System}
The 69-node distribution test system is a $12.66$ kV radial distribution system with $69$ nodes, $68$ branches and $5$ tie-lines. Therefore, the total number of branches in this system is $73$. All branches (including tie-lines) are numbered from $1$ to $73$ following our previous work \cite{gautam2020spanning}. The total load of the system is 3.80 MW. Detailed data of the system are given i \cite{savier2007impact}.
 \begin{figure}
    \centering
    \vspace{-1ex}
    \includegraphics[scale=0.55]{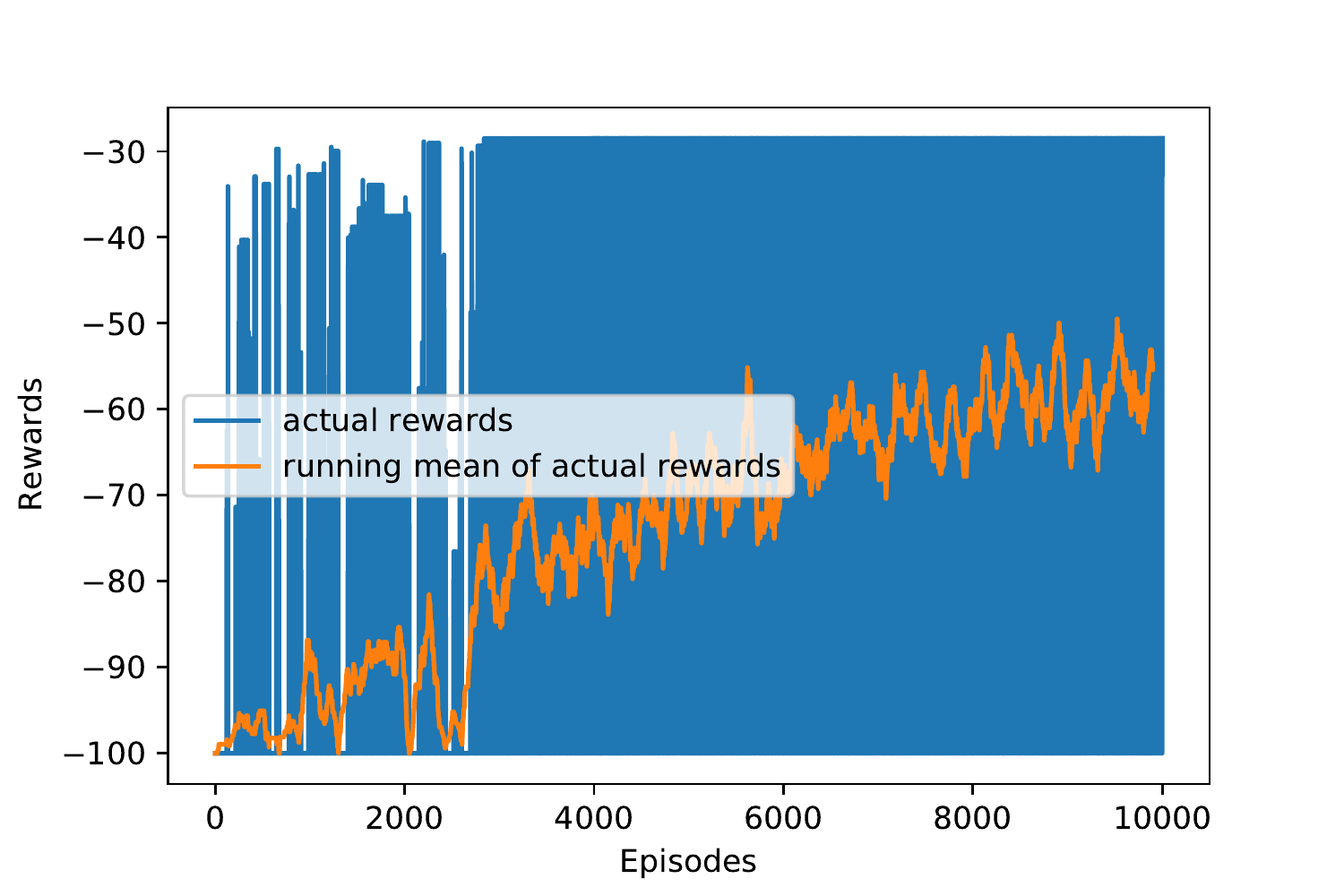}
    \vspace{-1.2ex}
    \caption{Rewards of training episodes for the 69-node system}
    \vspace{-1ex}
    \label{fig:69_rewards}
\end{figure}
\begin{figure}
    \centering
    \vspace{-1ex}
    \includegraphics[scale=0.55]{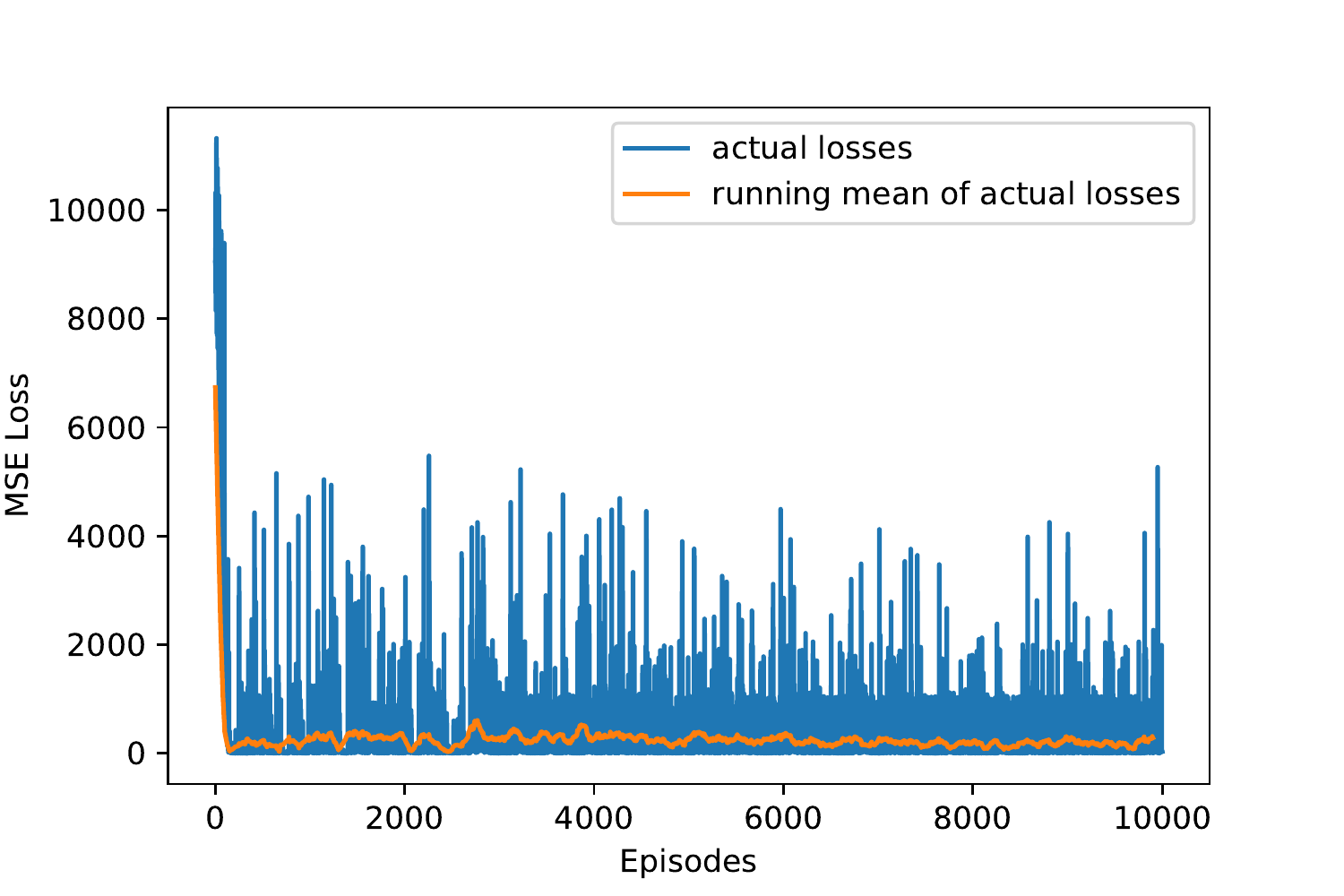}
    \caption{MSE of losses of training episodes for the 69-node system}
    \vspace{-1ex}
    \label{fig:69_losses}
\end{figure}
\begin{figure}
    \centering
    \vspace{-1ex}
    \includegraphics[scale=0.55]{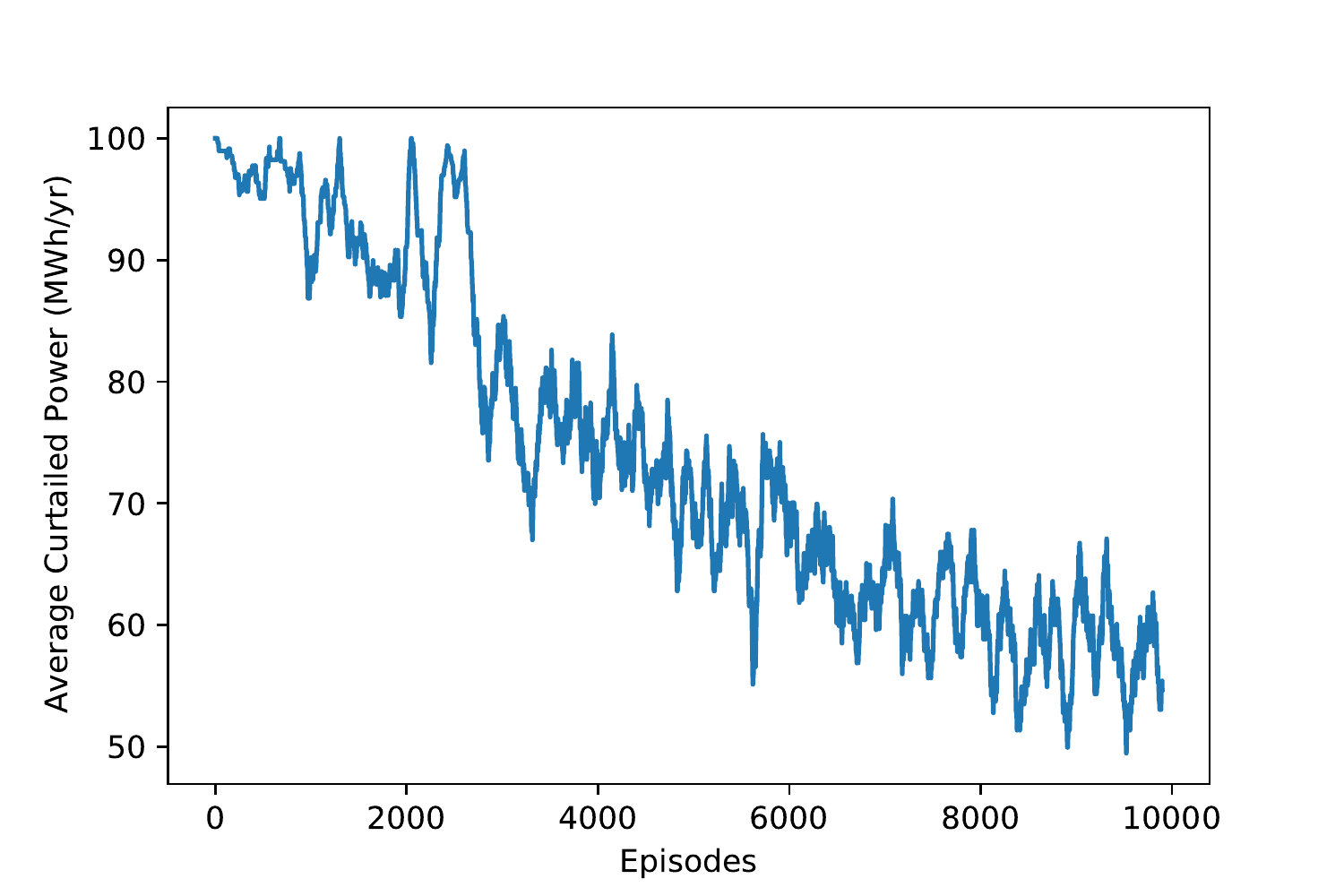}
    \vspace{-1.1ex}
    \caption{Average curtailed power of training episodes for the 69-node system}
    \vspace{-1ex}
    \label{fig:69_ENS}
\end{figure}

The training of the DQN for the 69-node system is performed for 10,000 episodes. The initial state of the system is the state with all tie-switches (i.e., edges 69, 70, 71, 72, and 73) are open. Initially, rewards are very low but then they increase as the number of episodes increases. Fig.~\ref{fig:69_rewards} shows the actual rewards and running mean (100-episode window) of actual rewards as the episode progresses. We can see from the Fig.~\ref{fig:69_rewards} that as the number of episode increases, the running mean of the rewards increases and saturates after nearly 8,000 episodes. Fig.~\ref{fig:69_losses} shows the actual values and running mean (100 episode window) of MSE losses as the episode progresses. Fig.~\ref{fig:69_losses} also shows that initially MSE loss is very high but its running mean remains low after a few episodes. Fig.~\ref{fig:69_ENS} shows the plot of running mean (100 episode window) of the average curtailed power as the episode progresses. From Fig.~\ref{fig:69_ENS} we can see that the the average curtailed power decreases as the number of episodes increases and settles down after nearly 8,000 episodes. The final value of the average curtailed power obtained during the training of the proposed framework is 28.48 MWh/yr. For this value of the average curtailed power, the open edges (branches) are 14, 18, 21, 58, and 69.

\section{Conclusion}\label{conclusion}
This paper has proposed a DRL-based framework to determine the configuration of a distribution network with the least value of the average curtailed power. During the training of the proposed framework, the weights of DNN were initialized with random values and the system state was initialized with the configuration with tie-switches open. The target Q values were computed based on reward the DRL agent gets from the environment. The predicted and target Q values were used to update the weights of DNN. Case studies were performed on 33-node and 69-node distribution test systems. The results exhibit the effectiveness of the proposed framework to improve the reliability level of distribution systems.

\section*{Acknowledgement}
This work was supported by the U.S. National Science Foundation (NSF) under Grant NSF 1847578. 

\bibliographystyle{IEEEtran}
\bibliography{References.bib}


\end{document}